\documentclass{mem}
\usepackage{natbib}\usepackage{txfonts}\usepackage{balance}
\usepackage{graphicx}
\usepackage[a4paper]{hyperref}
\idline{75}{282}
\begin{document}

\title{ Distances and Stellar Population properties using the SBF method }

\subtitle{}

\author{
M. \,Cantiello,
G. \, Raimondo,
E. \,Brocato
and I. \,Biscardi
}

\offprints{M. Cantiello}

\institute{ INAF -- Osservatorio Astronomico di Teramo, Via Maggini snc, I-64100 Teramo, Italy \\ \email{cantiello@oa-teramo.inaf.it} }

\authorrunning{M. Cantiello et al.}

\titlerunning{Distances and Stellar Populations with SBF}

\abstract{We present some results on the study of stellar population properties and distances of galaxies using the SBF technique. The applications
summarized here show that the Surface Brightness Fluctuations (SBF) method is able to $i)$ provide accurate distances of resolved and unresolved
stellar systems from $\sim$ 10~Kpc to $\sim$ 150 Mpc, and $ii)$ to reliably constrain the physical properties (e.g. age and metallicity) of
unresolved stellar systems.  } \maketitle{}

\section{Introduction}
The knowledge of the complex evolution of the stellar component in high redshift galaxies relies on how detailed is our understanding of the history
of stars in nearby galaxies. The best way to trace back the star formation episodes in a galaxy is to study resolved stars. However, this is possible
only for few nearby objects, tipically using the best observational facilities to date, and only for the brightest stars in the population.

Given such limitations several techniques have been proposed to disentangle the properties of unresolved stellar systems, one of these is the Surface
Brightness Fluctuations method \citep[SBF hereafter]{ts88}. The SBF technique was introduced as a distance indicator for elliptical galaxies within
$\sim$1-20 Mpc. After more than two decades of systematic applications it is now recognized that the SBF method works in a much larger distance
interval, and can be applied to a wider class of objects: ellipticals, bulges of spirals, dwarf ellipticals, globular clusters, etc. In addition, it
is well accepted that SBF magnitudes and colors represent a potential tool to analyze in details the physical and chemical properties of unresolved
and resolved stellar systems.

In this paper we will briefly discuss some applications of the SBF method done at the INAF-Observatory of Teramo by the SPoT\footnote{Teramo--Stellar
POpulations Tools group website: www.oa-teramo.inaf.it/SPoT} group.

\section{SBF by the SPoT Group}
\subsection{Models}
To derive distances from measured SBF magnitudes one needs a calibration of absolute SBF versus one integrated color \citep{tal90}. The calibration
is usually derived following an {\it empirical} approach \citep{tonry01}, which suffers for the usual drawbacks: one needs to know in advance the
distance of few objects in order to derive the zeropoint of the calibration; the calibration can be obtained for one filter at a time, etc. To avoid
these problems one can derive {\it theoretical} calibrations by means of stellar population models.

Taking advantage of the expertise by members of our group in the numerical sinthesys of stellar populations \citep[e.g.][]{brocato99,brocato00}, we
derived SBF models for stellar systems in the age range $\sim$50 Myr - 15 Gyr, for metallicity [Fe/H] from $\sim -2.3$ to 0.3 dex, for standard
UBVRIJHK photometric bands and HST filters (ACS, WFPC2 and NICMOS).

The reliability of the SBF models proposed by our group has been tested against various observational data \citep{c03,r05}.  As an example the
optical calibrations derived in V, I, and $z'$ using the SPoT models agree nicely with the empirical ones \citep[e.g.][]{biscardi08}. Also, the
comparison of near--IR models with available data has shown that the general behaviour of observations is well reproduced.

It is worth to emphasize that our approach to derive SBF models is original with respect to other methods. Our technique, in fact, is based on the
synthesis of Simple Stellar Populations (SSP) and it has two main advantages (see \citet{r05}, Sect. 3). First, multi-band Color-Magnitude Diagrams,
integrated colors and luminosity functions are available in addition to SBF models. Thus, models are tested against the observed SBF magnitudes {\it
and} also versus other astronomical observables of resolved and unresolved systems. Second, the SPoT code allows the user to set many different input
parameters (stellar tracks, mass loss, IMF, atmosphere models, etc.). As a consequence, the sensitivity of SBF to various properties of the stellar
population can be analyzed.

As an example, in our first release of SBF models we showed that the properties of hot evolved stars in old stellar systems (HB, Hot-HB, Post-AGB)
can be explored using SBF magnitudes in bands like B, or U. The future class of detectors with increased efficiency in the blue bands - like WFC3 on
board of HST, or the WSO satellite \citep{pagano07} - will be of great interest for SBF applications to the wavelength interval below 5000 $\AA$.

\begin{figure*}[t]
\centering \resizebox{6.5cm}{6.5cm}{\includegraphics[clip=true]{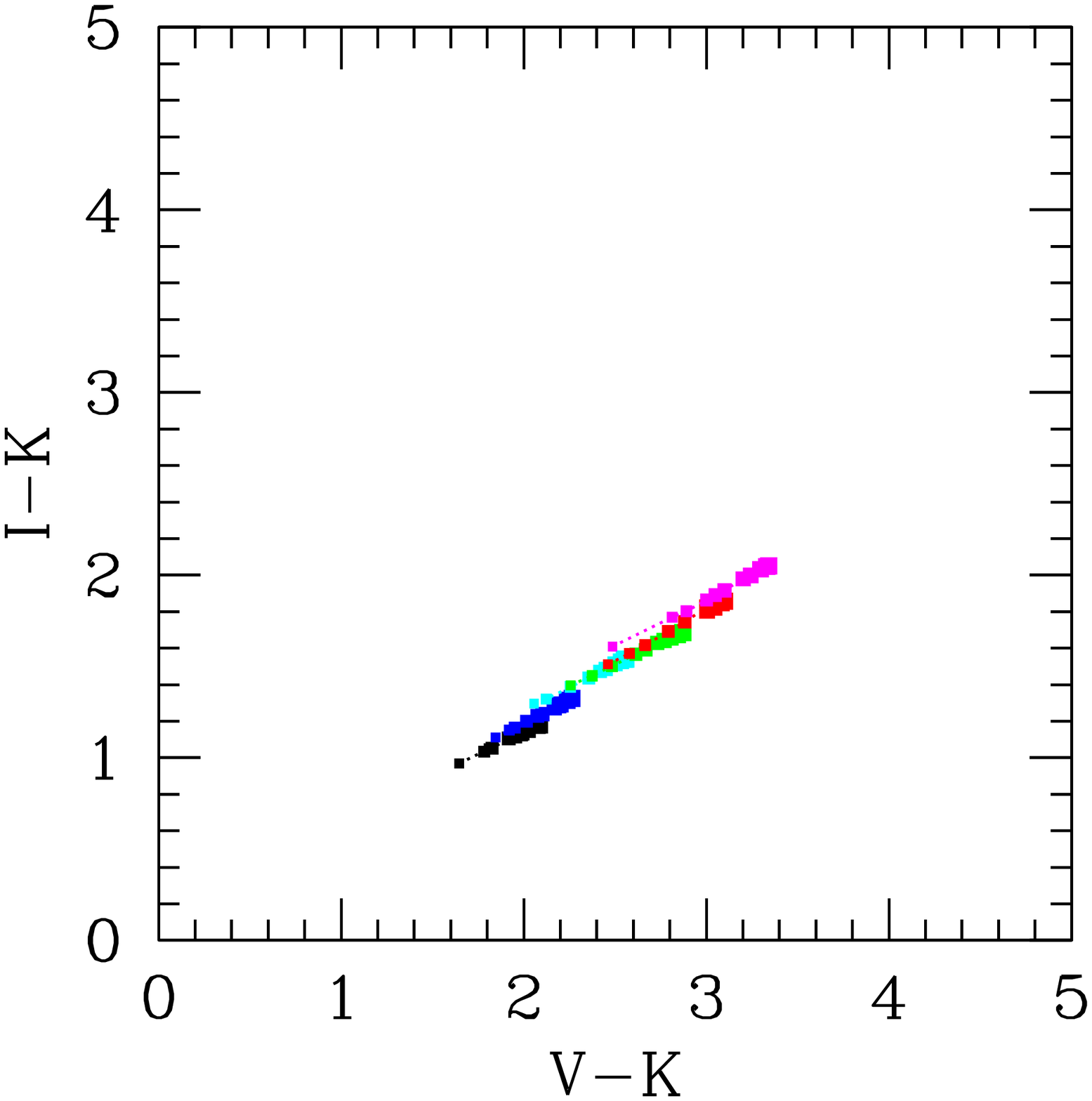}}
\resizebox{6.5cm}{6.5cm}{\includegraphics[clip=true]{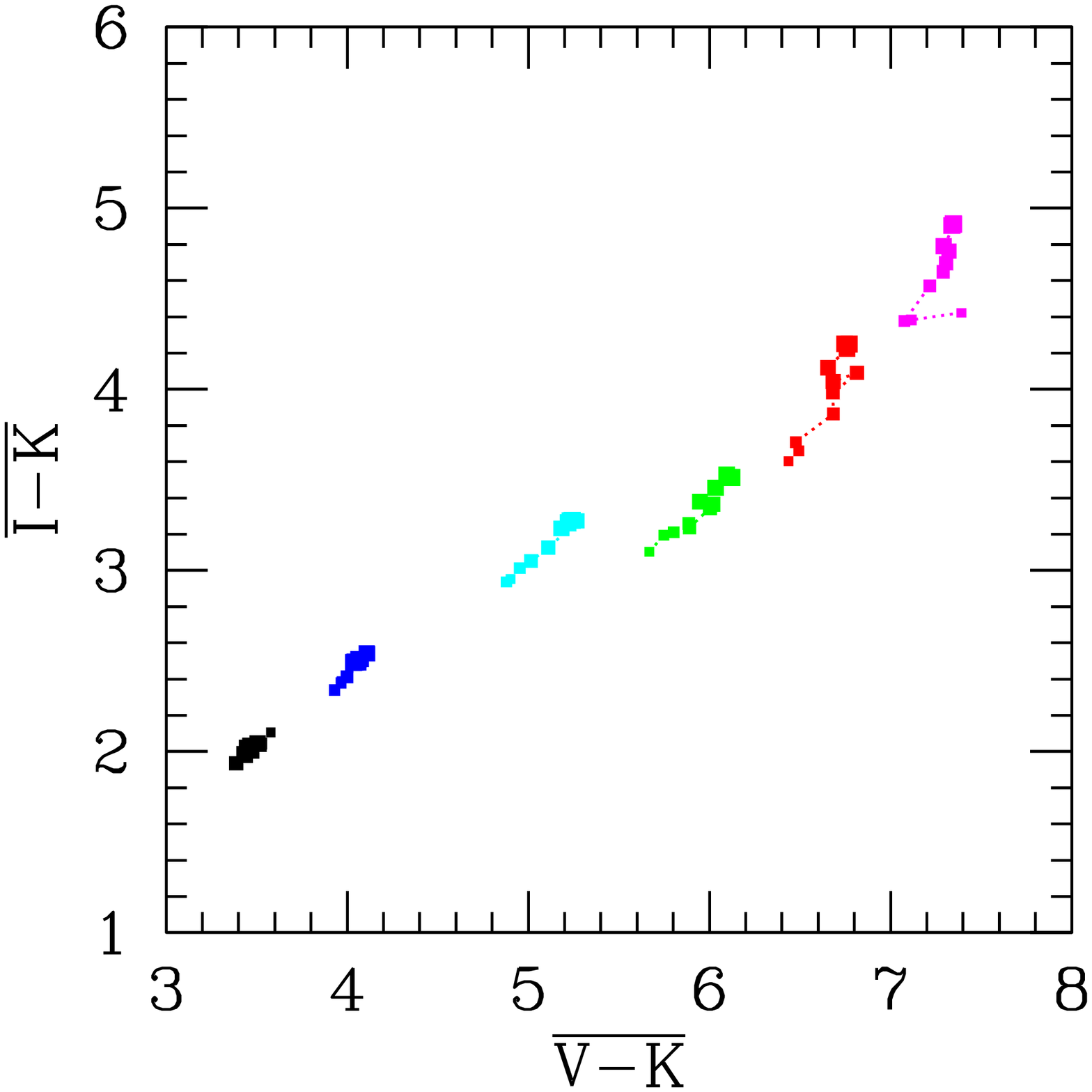}} \caption{\footnotesize Upper panel: Color-color diagram obtained using the
optical-to-near--IR {\it integrated} colors V-K and I-K. The metallicity is color coded - [Fe/H]=$-1.8$, $-1.3$, $-0.7$, $-0.3$, $0.0$, $+0.3$ dex
are shown in black, blue, cyan, green, red and magenta, respectively. For each metallicity ages between 1.5 and 14 Gyr are shown (larger symbols size
means older ages).  The SPoT models used are from \citet{r05}. Lower panel: As upper panel (same symbols, same models) except that SBF colors are
taken into account. Differently from upper panel, models with different chemical compositions are well separated from each other. Note that the
magnitude interval span by the x-- and y--axis is the same in both panels.} \label{colors}
\end{figure*}

\subsection{Measurements}

As mentioned above the SBF was proposed as a method to measure distances.  To estimate the distance of a galaxy the SBF magnitude and, usually, the
integrated $(V-I)_0$ color are obtained in one single annulus. However, it is reasonable to expect that SBF variations can be observed within a
single galaxy.

In 2003 we started a campaign aimed at the detection and study of SBF gradients in galaxies. For this purpose we have developed a procedure optimized
to reveal radial variations of SBF magnitudes. In our first study of SBF gradients - based on I-band images of eight ellipticals observed with ACS -
we found that the presence of SBF gradients seems to be correlated with the mass of the galaxy. In particular, less massive objects do not show
significant gradients \citep{c05}.

The presence of SBF gradients, though not unexpected, represents a further piece of information to study the evolutionary path of galaxies. Moreover,
gradients do not depend on the distance of the object, so any comparison with models is free from this heavy uncertainty. Comparing data with SPoT
models we have pointed out that the amplitude of the SBF versus color gradient seems to be dominated by a [Fe/H] variation along the radius of the
galaxy, rather than to age variations. Such result appears even more meaningful if one considers that the only galaxy in our sample with a gradient
likely dominated by age variations is NGC 1344, a galaxy that shows morphological irregularities possibly related to a recent gravitational
interaction.

This study has been recently extended, again using ACS data, both in the I- and V-band \citep{c07}. The study comprises 14 galaxies spanning a large
interval of total magnitudes ($\sim$10 mag), and with very different properties. The new data confirm the results found in our previous study, and
suggest that a larger database of SBF measurements will provide a valuable tool to analyze stars in galaxies.

In general, our study on the use of SBF as a tracer of stellar population properties - via SBF gradients, absolute SBF magnitudes, or optical SBF
colors - demonstrated that the metallicity of the dominant stellar component can be better confined with respect to age, unless optical to near--IR
SBF magnitudes are coupled (see next section).

Besides the exploration of the properties of unresolved stellar populations, we have also taken into account the chance to measure SBF for the study
of resolved stellar systems, and galaxies at large distances.  In the first case we have carried out a dedicated study on star clusters in the
Magellanic Clouds, providing the first optical SBF measurements for different MC star clusters, and analysing in details their optical and near--IR
SBF properties through data to models comparisons \citep{r05}. More recently, we measured SBF magnitudes for four distant ellipticals observed with
ACS. The large distances of the objects observed allowed us to estimate H$_0$ - see Biscardi et al.'s article in this volume, and references therein.

\subsection{SBF--colors }

To date the study of unresolved stellar populations using SBF has been carried out using three different approaches: $i$) absolute SBF magnitudes;
$ii$) optical SBF colors; $iii$) SBF gradients. However, since the main application of SBF is to derive distances, none of the applications existing
in literature is optimized for stellar population studies.

In Fig.~\ref{colors} we show two color-color panels, both obtained using the SPoT models for ages between 1.5 and 14 Gyr, and metallicity [Fe/H] from
$-1.8$ to 0.3 dex. The upper panel of the figure shows the integrated optical-to-near--IR colors I-K and V-K, the lower panel uses the same I-K and
V-K colors but SBF magnitudes are taken into account.  It is easy to recognize in the upper panel that the age-metallicity degeneracy strongly
affects models. On the contrary, the lower panel shows that optical-to-near--IR SBF color models in the age and metallicity regimes considered are
well separated, and a data to models comparison will help to put substantial constraints to the metallicity of the system. At the same time, thanks
to the large separation between models at fixed [Fe/H], these SBF-color panels can also be used to constrain the age of the dominant stellar
component within an interval better confined with respect to, e.g., integrated colors.

The applications of the SBF-color technique described above are to date limited to optical colors, or refer to inhomogeneous sets of optical and
near--IR SBF measurements \citep[e.g.][]{jensen03}. In all present applications, however, the comparison of data with models confirm the reliability
of the technique proposed and of models, and supports the potential of SBF colors to explore the physical and chemical properties of unresolved
stellar populations.

\section{Conclusions and Future perspectives}

The SBF technique is to date one of the most reliable distance indicators for elliptical galaxies. However, to our point of view, this technique is
underestimated with respect to its real potentiality.

Concerning distance measurements, the SBF method is able to provide distances from few Kpc up to $\sim$150 Mpc with present observing facilities, and
possibly to much larger distances with future instrumentations. Such unique characteristic gives the SBF the potential to cover the distance scale
ladder from local to low redshift ($z\leq0.05$) distances. Thus, SBF measurements, coupled with a reliable calibration of absolute SBF magnitudes,
provide a great opportunity to substantially reduce the systematic uncertainty that affects the cosmological distance scale.

With regard to stellar population analysis, there are only few studies dedicated to this topic based on the SBF method. However, it is now evident
that SBF can greatly improve our understanding of the properties of unresolved stellar systems. We presented SPoT models in a specific SBF
optical-to-near--IR color plane showing the potential of SBF colors to substantially remove the age-metallicity degeneracy. Future application of
this technique - possibly coupled with the measure SBF-color gradients attainable with the next generation optical and near--IR large FoV detectors -
will provide significant constraints to the knowledge on the formation and evolution of the stellar component in low-redshift galaxies.

\begin{acknowledgements}
It is a pleasure to aknowledge J. P. Blakeslee and S. Mei for their helpful contribution to the topic of SBF and SBF gradients measurements.
\end{acknowledgements}

\bibliographystyle{aa}

\end{document}